\documentclass[aps,showpacs,prd]{revtex4-1}

\usepackage{graphicx,hyperref}
\usepackage{amsmath,amssymb}

\newcommand{\bra}[1]{\left\langle#1\phantom{.}\!\right|}
\newcommand{\ket}[1]{\left|\phantom{.}\!#1\right\rangle}
\newcommand{\matens}[3]{\langle#1|#2|#3\rangle}
\newcommand{\rket}[1]{|#1)}
\newcommand{\rbra}[1]{(#1|}
\newcommand{\inp}[2]{\langle#1|#2\rangle}
\newcommand{\rinp}[2]{(#1|#2)}
\newcommand{\psih}{\hat{\psi}}

\newcommand{\mycomment}[1]{}

\newcommand{\ncp}{\lambda} 
\newcommand{\hs}[1]{\hspace{#1cm}}

\newcommand{\aop}[1]{\hat{a}_{#1}}
\newcommand{\adop}[1]{\hat{a}^\dag_{#1}}
\newcommand{\HQ}{\mathcal{H}_Q}
\newcommand{\HC}{\mathcal{H}_C}
\newcommand{\caly}{\hat{\mathcal{Y}}}
\newcommand{\kappab}{\bar{\kappa}}
\newcommand{\matcc}[2]{\begin{array}{ll} #1 \\ #2 \end{array}}
\newcommand{\gbar}{\bar{g}}
\newcommand{\zbar}{z^*}
\newcommand{\nhat}{\hat{n}}

\begin{document}

\title{Scattering in three-dimensional fuzzy space}

\author{J~N~Kriel$^1$, H~W~Groenewald$^1$ and F~G~Scholtz$^{1,2}$}
\affiliation{$^1$Institute of Theoretical Physics, University of Stellenbosch, Stellenbosch 7600, South Africa}
\affiliation{$^2$National Institute for Theoretical Physics (NITheP), Stellenbosch 7600, South Africa}

\begin{abstract}

We develop scattering theory in a non-commutative space defined by a $su(2)$ coordinate algebra. By introducing a positive operator valued measure as a replacement for strong position measurements, we are able to derive explicit expressions for the probability current, differential and total cross-sections. We show that at low incident energies the kinematics of these expressions is identical to that of commutative scattering theory. The consequences of spacial non-commutativity are found to be more pronounced at the dynamical level where, even at low incident energies, the phase shifts of the partial waves can deviate strongly from commutative results. This is demonstrated for scattering from a spherical well. The impact of non-commutativity on the well's spectrum and on the properties of its bound and scattering states are considered in detail. It is found that for sufficiently large well-depths the potential effectively becomes repulsive and that the cross-section tends towards that of hard sphere scattering. This can occur even at low incident energies when the particle's wave-length inside the well becomes comparable to the non-commutative length-scale.

\end{abstract}
\pacs{11.10.Nx}
\vspace{2pc}

\maketitle
\section{Introduction}
\label{intro}

The description of space-time at short length-scales has emerged as one of the central problems in modern physics \cite{seiberg_emergent_2006}. While there is growing consensus that our notion of space-time at the Planck length requires major revision, there seems to be much less agreement on what form an appropriate description at these scales should take. One possibility is that of non-commutative space-time. This scenario was originally proposed by Snyder \cite{snyder_quantized_1947} in an attempt to avoid the infinities encountered in quantum field theories. More recently, interest in these ideas have been stimulated by the arguments of Doplicher et al. \cite{doplicher_quantum_1995} and the emergence of non-commutative coordinates in the low energy limit of certain string theories \cite{seiberg_string_1999}.\\

The simplest version of non-trivial space-time commutation relations is $\left[\hat{x}_\mu , \hat{x}_\nu \right] = i \theta_{\mu \nu}$ with $\theta_{\mu \nu}$ a set of scalar constants. This case has been studied extensively, but still presents a number of difficulties. In particular, these commutation relations break rotational invariance in $3$-dimensional quantum theories and Lorentz invariance in $3+1$-dimensional quantum field theories.  In addition, these field theories have the undesirable property of UV/IR mixing, which may jeopardise their perturbative renormalisability \cite{douglas_noncommutative_2001}. In the case of 3+1-dimensional quantum field theories, it was recently shown that the Lorentz symmetry is restored upon twisting \cite{chaichian_new_2005}. Despite this insight several outstanding and controversial issues still plague the twisted implementation of the Lorentz symmetry \cite{pinzul_uv/ir_2012}. The first difficulty one encounters is to carry out the standard Noether analysis and to identify conserved charges for the twisted Lorentz symmetry. Another obstacle is the quantisation of these theories, where one can adopt the standard quantisation procedure or, in addition, also deform the canonical commutation relations. On the level of the functional integral this amounts to altering the measure. This has rather drastic consequences such as the absence of UV/IR mixing. Indeed, in \cite{pinzul_uv/ir_2012} it is argued that UV/IR mixing may be related to a quantum anomaly of the Lorentz group, which is very closely related to the choice of functional integral measure. At this point there seems to be no consensus between these different points of view. See \cite{pinzul_uv/ir_2012} and references therein for further discussions.\\

A more direct way to overcome the breaking of rotational symmetry, at least in 3-dimensional quantum mechanics, is to adopt fuzzy sphere or $su(2)$ commutation relations \cite{galikova_coulomb_2013-1,galikova_hydrogen_2012,chandra_spectrum_2014} of the form
\begin{equation}
[\hat{x}_i,\hat{x}_j] = 2i\ncp\epsilon_{ijk}\hat{x}_k
\end{equation}
where $\ncp$ is the non-commutative length parameter. The Coulomb problem \cite{galikova_coulomb_2013-1,galikova_hydrogen_2012} and spherical well \cite{chandra_spectrum_2014} have been investigated in this framework. The latter results were also used to investigate the thermodynamics of non-commutative Fermi gases \cite{scholtz_thermodynamics_2015} which was found to deviate strongly from that of the commutative case at high densities and temperatures. The latter effects may have observational consequences for very dense astro-physical objects \cite{scholtz_thermodynamics_2015}.\\

In the search for possible signatures of non-commutativity the most directly measurable quantity available to us are scattering cross-sections. It would therefore be useful to establish a formalism for scattering in non-commutative space in order to identify possible systematic trends that could serve as indicators of non-commutativity. Scattering theory in two dimensional non-commutative space has been developed in detail in \cite{thom_bound_2009} and an analysis of the S-matrix for the Coulomb potential in three dimensions was performed in \cite{galikova_coulomb_2013,galikova_coulomb_2013-1}. However, a detailed development of scattering theory and the phase shift analysis in three dimensional fuzzy space is still lacking. This is the purpose of this paper.\\

The paper is organised as follows: In section \ref{sec:formalism} we review the formalism of quantum mechanics in fuzzy space before moving on to a discussion of the position representation, continuity equation and probability current. The solutions to the free particle problem are discussed and expressions for their probability currents are derived. Section \ref{sec:scattering-formalism} develops non-commutative scattering theory via the partial wave expansion culminating in explicit expressions for the differential and total scattering cross-sections. Section \ref{sec:well-spectrum} summarises the relevant aspects of the spectrum and eigenstates of the attractive spherical well potential which are then used in the development and interpretation of the scattering analysis in section \ref{sec:well-scattering}. Finally, section \ref{sec:conclusions} summarises the results and draws conclusions. 
\section{Formalism}
\label{sec:formalism}
\subsection{Quantum Mechanics in Fuzzy Space}
We begin with a brief account of quantum mechanics in the non-commutative three-dimensional space \cite{galikova_coulomb_2013-1} characterised by the rotationally invariant $su(2)$ coordinate algebra
\begin{equation}
	[\hat{x}_i,\hat{x}_j]=2i\ncp \varepsilon_{ijk}\hat{x}_k.
\label{eq:nc-coor-coms}
\end{equation}
Here $\ncp$ is the non-commutative length parameter. The Casimir operator $\hat{x}^2=\hat{x}_i\hat{x}_i$ is associated with the square of the radial distance and its eigenvalues are determined by the $su(2)$ representation under consideration. The first step towards formulating quantum mechanics on this space is to construct a representation of the coordinate algebra on a Hilbert space $\HC$, referred to as the classical configuration space. The idea is to mimic $R^3$ as a collection of concentric shells (an onion structure) where the radii of the shells are quantised according to the eigenvalues of  $\hat{x}^2$. This requires that $\HC$ carries a single copy of each $su(2)$ irrep. A concrete realisation of this setup is provided by the Schwinger construction, which utilises two sets of boson creation and annihilation operators to build the desired representation of $su(2)$. These operators satisfy
\begin{equation}
	[\hat{a}_\alpha,\hat{a}^\dagger_\beta]=\delta_{\alpha\beta}\quad\mathrm{and}\quad[\hat{a}_\alpha,\hat{a}_\beta] = [\hat{a}^\dagger_\alpha,\hat{a}^\dagger_\beta]=0\quad\mathrm{with}\quad\alpha,\beta=1,2
\label{eq:operator-coms}
\end{equation}
and we identify the corresponding Fock space with $\HC={\rm span}\{\ket{n_1,n_2}:\adop{i}\aop{i}\ket{n_1,n_2}=n_i\ket{n_1,n_2}\}$. The coordinates themselves are realised as
\begin{equation}
	\hat{x}_i = \ncp\hat{a}^\dagger_\alpha\sigma^{(i)}_{\alpha\beta}\hat{a}_\beta
\label{eq:nc-coordinates}
\end{equation}
where $\{\sigma^{(i)}\}$ are the Pauli spin matrices. The Casimir operator now reads $\hat{x}^2=\hat{x}_i\hat{x}_i=\ncp^2\nhat (\nhat+2)$ with $\nhat=\hat{a}_1^\dagger\hat{a}_1+\hat{a}_2^\dagger\hat{a}_2$ and so it is clear that each $su(2)$ irrep occurs exactly once in $\HC$. As a measure of radial distance we will use
\begin{equation}
	\hat{r} = \ncp(\nhat+1)
\label{eq:distance-op}
\end{equation}
which is, up to order $\mathcal{O}(\ncp^2)$, the square root of $\hat{x}^2$. The quantum Hilbert space $\HQ$ is now defined as the algebra of operators generated by the coordinates, i.e. the operators acting on $\HC$ that commute with $\hat{x}^2$ and have a finite norm with respect to a weighted Hilbert-Schmidt inner product \cite{galikova_coulomb_2013-1}: 
\begin{eqnarray}
\label{eq:defineHQ}
{\cal H}_Q=\left\{\hat{\psi}=\hs{-0.3}\sum_{m_i,n_i = 0}^{\infty}\hs{-0.2}C^{m_1,m_2}_{n_1,n_2} (\adop{1})^{m_1}(\adop{2})^{m_2}\aop{1}^{n_1}\aop{2}^{n_2}: m_1+m_2=n_1+n_2\ \ {\rm and}\ \ {\rm tr_C}(\hat{\psi}^\dagger\hat{r}\hat{\psi})<\infty\right\}.
\end{eqnarray}
The inner product on $\HQ$ is
\begin{equation}
	\rinp{\hat{\psi}}{\hat{\phi}}=4\pi\ncp^2{\rm tr}_C(\hat{\psi}^\dag\hat{r}\hat{\phi})
\end{equation}
with the trace taken over $\HC$. We will use the standard $\ket{\cdot}$ notation for elements of $\HC$ and, where appropriate, $\rket{\cdot}$ for elements of $\HQ$. Quantum observables are identified with self-adjoint operators acting on $\HQ$. These include the coordinates which act through left multiplication as
\begin{equation}
	\hat{X}_i\psih=\hat{x}_i\psih
\end{equation}
and the angular momentum operators which act adjointly according to
\begin{equation}
	\hat{L}_i\psih = \frac{\hbar}{2\ncp}[\hat{x}_i, \psih]\quad{\rm with}\quad[\hat{L}_i,\hat{L}_j] = i\hbar\varepsilon_{ijk}\hat{L}_k.
\label{eq:nc-angular-ops}
\end{equation}
The non-commutative analogue of the Laplacian is defined as
\begin{equation}
 \hat{\Delta}_\ncp\psih=-\frac{1}{\ncp\hat{r}}[\hat{a}^\dagger_\alpha,[\hat{a}_\alpha,\psih]]
\label{eq:nc-laplacian}
\end{equation}
and can be shown to commute with the three angular momentum operators \cite{galikova_coulomb_2013-1}.
\subsection{Position Representation}
To develop a formalism of scattering in non-commutative space we require a means of assigning a spacial probability density and corresponding probability current to elements of the quantum Hilbert space $\HQ$. This task is complicated by the fact that the spacial coordinates no longer commute, as this excludes the standard construction via position eigenstates. We will therefore adopt the approach of \cite{scholtz_formulation_2009} and introduce a positive operator valued measure (POVM) based on minimum uncertainty states in $\HC$. In POVM language this amounts to replacing the notion of a strong position measurement by a weak measurement.\\

Recall that $\HC$ is a bosonic Fock space carrying a reducible representation of the $su(2)$ coordinate algebra. This can be made explicit by labelling the basis states as $\ket{j,m}\equiv\ket{n_1,n_2}$ with $j=(n_1+n_2)/2=n/2$ and $m=(n_1-n_2)/2$. We can regard the state $\ket{j,m}$ as representing a particle localised at a radial distance of $r=\ncp(2j+1)$, but delocalised in the two angular directions. We wish to identify those states for which this delocalisation is at a minimum. For a given value of $j$ and a unit vector $\hat{u}=(\sin(\theta)\cos(\theta),\sin(\theta)\sin(\phi),\cos(\theta))$ we seek the state for which the component of $\hat{\vec{x}}=(\hat{x}_1,\hat{x}_2,\hat{x}_3)$ along $\hat{u}$ is maximal. This would ensure the best possible localisation of the particle at $r=\ncp(2j+1)$ around the point $(\theta,\phi)$. These are precisely the $su(2)$ coherent states \cite{perelomov_generalized_1986} defined by
\begin{equation}
	\ket{n,z}=\frac{1}{(1+z\zbar)^{n/2}}e^{z\hat{x}_+}\ket{j=\frac{n}{2},m=-\frac{n}{2}}
\end{equation}
where $\hat{x}_+=\hat{x}_1+\hat{x}_2$ and $z\in\mathbb{C}$. These states satisfy $(\hat{u}\cdot\hat{\vec{x}})\ket{n,z}=\ncp n\ket{n,z}$ where the link between $z$ and the angular variables is $z=\cot(\theta/2)e^{-i\phi}$. The identity on $\HC$ can now be resolved as
\begin{equation}
	\hat{I}_C=\sum_{n=0}^\infty\int dzd\zbar  \,\mu_n(z,\zbar )\ket{n,z}\bra{n,z}\hs{1}{\rm with}\hs{1}\mu_n(z,\zbar )=\frac{n+1}{\pi(1+z\zbar )^2}.
\end{equation}
A similar construction is possible on $\HQ$ using the states
\begin{equation}
	\rket{n,z,w}=\frac{\ket{n,z}\bra{n,w}}{\sqrt{4\pi\ncp^3(n+1)}}
\end{equation}
in terms of which 
\begin{equation}
		\hat{I}_Q=\sum_{n=0}^\infty\int dzd\zbar \,\mu_n(z,\zbar )\int dw dw^*\,\mu_n(w,w^*)\rket{n,z,w}\rbra{n,z,w}.
\end{equation}
Following \cite{alexanian_generalized_2001}, this can be recast in the form
\begin{equation}
\hat{I}_Q=\sum_{n=0}^\infty\int d\zbar  dz\mu_n(z,\zbar ) |n,z,z)\ast_n (n,z,z|
\end{equation}
where the star product is defined by 
\begin{equation}
\ast_n= \int dw dw^*\,e^{\stackrel{\leftarrow}{\partial}_{z}w}\mu(z+w,\zbar+w^*)|\inp{n,z}{n,z+w}|^2 e^{w^*\stackrel{\rightarrow}{\partial}_{\zbar }}.
\label{eq:star-product}
\end{equation}
It follows that the operators
\begin{equation}
\hat{\pi}_{n,z}=|n,z,z)\ast_n (n,z,z|
\end{equation}
constitute a positive operator valued measure. For a particle described by the pure state density matrix $\rho=|\psih)(\psih|$ the probability density at a radial distance $r=\ncp(n+1)$ and angular variables $z=z(\theta,\phi)$ is then given by 
\begin{equation}
P(n,z)={\rm tr_Q}(\hat{\pi}_{n,z}\rho)=(\psih|n,z,z)\ast_n(n,z,z|\psih)
\end{equation} 
where the position representation of $\psih$ is
\begin{equation}
(n,z,z|\psih)=\sqrt{4\pi\ncp^3(n+1)}\langle n,z|\psih|n,z\rangle.
\label{eq:position-rep}
\end{equation}
The matrix element $\langle n,z|\psih|n,z\rangle$ is often referred to as the symbol of the operator $\psih$. Together with the star-product we have the useful identity \cite{alexanian_generalized_2001}
\begin{equation}
	\langle n,z|\hat{\psi}\hat{\phi}|n,z\rangle=\langle n,z|\hat{\psi}|n,z\rangle\ast_n\langle n,z|\hat{\phi}|n,z\rangle
	\label{eq:starproductfactorise}
\end{equation}
 which allows for the compact representation of the probability density as
\begin{equation}
P(n,z)=4\pi\ncp^3(n+1)\langle n,z|\hat{\psi}^\dag\hat{\psi}|n,z\rangle.
\end{equation}
\subsection{Continuity Equation}
To identify the probability current we will consider the continuity equation for the probability density $P(n,z,t)$ where the time evolution is generated by the Hamiltonian
\begin{equation}
	\hat{H}_\ncp=-\frac{\hbar^2}{2m_0}\hat{\Delta}_\ncp+V(\hat{r}).
\end{equation}
It follows that
\begin{align}
	\frac{d}{dt}\left[\mu_n(z,\zbar )P(n,z,t)\right]&=\frac{4\ncp^3}{i\hbar}\frac{(n+1)^2}{(1+z\zbar )^2}\matens{n,z}{\psih^\dag\hat{H}\psih-(\hat{H}\psih)^\dag\psih}{n,z}\\
	&=\frac{2i\hbar\ncp}{m_0}\frac{n+1}{(1+z\zbar )^2}\matens{n,z}{[\psih^\dag\adop{\alpha}\psih,\aop{\alpha}]-[\adop{\alpha},\psih^\dag\aop{\alpha}\psih]}{n,z}
	\label{eq:continuity1}
\end{align}
where $V(\hat{r})$ has fallen away as it commutes with $\psih$. We have also included the measure associated with the POVM $\hat{\pi}_{n,z}$ explicitly. The task is now to express the right hand side above as a divergence in spherical coordinates, albeit one where the radial direction is discretized. To this end the following identities, in which $\ket{n,z}_u\equiv(1+z\zbar )^{n/2}\ket{n,z}$ is the unnormalised coherent state, are very useful
\begin{align}
	\aop{1}\ket{n,z}_u&=\sqrt{n}z\ket{n-1,z}_u&\hs{1}\aop{2}\ket{n,z}_u&=\sqrt{n}\ket{n-1,z}_u \\
	\adop{1}\ket{n,z}_u&=\sqrt{n+1}\partial_z\ket{n+1,z}_u&\hs{1}\adop{2}\ket{n,z}_u&=\sqrt{n+1}(n+1-z\partial_z)\ket{n+1,z}_u
	\label{eq:a-actions}
\end{align}
It is now possible to bring \eqref{eq:continuity1} into the more recognisable form
\begin{align}
	\frac{d}{dt}\left[\mu_n(z,\zbar )P(n,z,t)\right]&=-\left[\sin(\theta)\Delta_r[r(r+\ncp) j_r(r,\theta,\phi)]+r\frac{\partial}{\partial \theta}[\sin(\theta)j_\theta(r,\theta,\phi)]+r\frac{\partial}{\partial \phi}j_\phi(r,\theta,\phi)\right]
\end{align}
where $r=\ncp(n+1)$ and $\Delta_r$ is the discrete radial derivative acting as $\Delta_r[f(r)]\equiv\frac{f(r)-f(r-\ncp)}{\ncp}$. The right-hand side is indeed a total divergence which includes the Jacobian $J=r^2\sin(\theta)$. The spherical components of the current are found to be
\begin{equation}
	j_r(r,\theta,\phi)=-\frac{\hbar}{m_0}\frac{{\rm Im}\left[M_1+M_2\right]}{n+1}\hs{1}{\rm and}\hs{1}j_{\theta/\phi}(r,\theta,\phi)=\frac{\hbar}{m_0}\frac{{\rm Im}/{\rm Re}\left[M_1-z\zbar M_2\right]}{\sqrt{zz^*}(n+1)}\label{eq:current-components}
\end{equation}
with $M_{1,2}$ the matrix elements $M_\alpha=\matens{n+1,z}{\psih^\dag\adop{\alpha}\psih\aop{\alpha}}{n+1,z}$. As before the angular dependence in \eqref{eq:current-components} enters through $z=\cot(\theta/2)e^{-i\phi}$.
\subsection{Free Particle Solutions}
The solutions of the free particle problem will play a crucial role in the general scattering formalism and the treatment of the spherical well. In this section and the next we collect the relevant results. The free particle Schr\"odinger equation reads
\begin{equation}
	\hat{H}_\ncp\psih=-\frac{\hbar^2}{2m_0}\hat{\Delta}_\ncp\psih=E\psih
	\label{eq:FPSE}
\end{equation}
and admits both plane-wave and radial solutions. A natural candidate for the former is $\psih_{\vec{k}}=e^{i\vec{k}\cdot\hat{\vec{x}}}$ which transforms under rotations as $\exp[-i\phi\hat{u}\cdot\hat{\vec{L}}/\hbar]\psih_{\vec{k}}=\psih_{R\vec{k}}$ with $R\equiv R_\phi(\hat{u})$ the corresponding rotation matrix. Due to the rotational invariance of the Schr\"odinger equation we can therefore focus on the case where $\vec{k}$ is orientated in the $z$-direction. We will denote the magnitude of $\vec{k}$ by $\bar{k}$ and reserve the symbol $k$ for later use. The state now reads
\begin{equation}
	\psih_{\bar{k}\hat{z}}=e^{i\bar{k}\hat{x}_3}=e^{i\bar{k}\ncp(\adop{1}\aop{1}-\adop{2}\aop{2})}
	\label{eq:kplanewave}
\end{equation}
and it is straightforward to check that 
\begin{equation}
	\hat{H}_\ncp\psih_{\bar{k}\hat{z}}=\frac{2\hbar^2}{m_0\ncp^2}\sin^2\left(\frac{\bar{k}\ncp}{2}\right)\psih_{\bar{k}\hat{z}}.
\end{equation}
Non-commutativity has clearly modified the dispersion relation at high energies and introduced an energy upper-bound of $E_{max}=\frac{2\hbar^2}{m_0\ncp^2}$ \cite{galikova_coulomb_2013-1,chandra_spectrum_2014}. This coincides with the observation that for the states in \eqref{eq:kplanewave} to be linearly independent the value of $\bar{k}$ must be restricted to the interval $\bar{k}\in\left[0,\frac{\pi}{\ncp}\right)$. We will use the notation
\begin{equation}
	E=\frac{\hbar^2\kappa^2}{2m_0\ncp^2}\hs{1}{\rm with}\hs{1}\kappa=2\sin\left(\frac{\kappab}{2}\right)\hs{1}{\rm and}\hs{1}\kappab=\bar{k}\ncp.
	\label{eq:planewaveenergy}
\end{equation}
Also, we define $k$ such that $\kappa=k\ncp$ as this allows for the more familiar looking expression $E=\frac{\hbar^2k^2}{2m_0}$.\\

Next we consider free particle angular momentum eigenstates. These take the form \cite{galikova_coulomb_2013-1}
\begin{equation}
	\psih_{lm}=\sum_{(lm)}\frac{(\hat{a}^\dagger_1)^{m_1}(\hat{a}^\dagger_2)^{m_2}}{m_1!\,m_2!}\gbar(\nhat,\kappa)\frac{(\hat{a}_1)^{n_1}(-\hat{a}_2)^{n_2}}{n_1!\,n_2!}
\label{eq:standard-eigenfunctions}
\end{equation}
with $l=0,1,2,\ldots$ and $m=-l,\ldots,+l$. The summation above runs over the non-negative integers ($m_1,m_2,n_1,n_2$) satisfying $m_1+m_2=n_1+n_2=l$ and $m_1-m_2-n_1+n_2=2m$. Inserting $\psih_{lm}$ into the Schr\"odinger equation yields a difference equation for $\gbar$ which admits two linearly independent solutions
\begin{align}
	\gbar_{J,l}(n,\kappa)&=\left[\frac{\sqrt{\pi}\sin^{l+1}(\kappab)}{2^{l+1}\Gamma[\frac{3}{2}+\ell]}\right]\cos^n(\kappab)\,{}_2 F_1\left[\frac{1-n}{2},-\frac{n}{2},\frac{3}{2}+\ell,-\tan^2(\kappab)\right] & (n\geq 0)\\
	\gbar_{Y,l}(n,\kappa)&=\left[-\frac{\sqrt{\pi}(-2)^l\cos^{l+1}(\kappab)}{\tan^l(\kappab)\Gamma[\frac{1}{2}-\ell]}\right]\frac{n!\cos^n(\kappab)}{\Gamma[2+2l+n]}\,{}_2 F_1\left[\frac{-1-l-n}{2},-l-\frac{n}{2},\frac{1}{2}-\ell,-\tan^2(\kappab)\right]& (n>0)
\end{align}
These are the non-commutative analogues of the spherical Bessel and Neumann functions. See section 7 of \cite{chandra_spectrum_2014} for a more detailed derivation. Note that $\gbar_{Y,l}$ fails to solve the Schr\"odinger equation at $n=0$, i.e. at the origin. As such it plays no role in the free particle problem, but will enter in the scattering formalism later. It is also convenient to define the analogue of the Hankel functions of the first kind as $\gbar_{H,l}=\gbar_{J,l}+i\gbar_{Y,l}$.\\

We note that \eqref{eq:standard-eigenfunctions} can also be written as
\begin{equation}
	\psih_{lm}=\gbar(\nhat-l,\kappa)\caly_{lm}\hs{1}{\rm with}\hs{1}\caly_{lm}=\sum_{(lm)}\frac{(\hat{a}^\dagger_1)^{m_1}(\hat{a}^\dagger_2)^{m_2}}{m_1!\,m_2!}\frac{(\hat{a}_1)^{n_1}(-\hat{a}_2)^{n_2}}{n_1!\,n_2!}.
\label{eq:standard-eigenfunctions2}
\end{equation}
The latter operator plays the role of a spherical harmonic and has the usual eigenvalues with respect to $\hat{L}_3$ and $\hat{L}^2$. This link can be made explicit by considering the symbol of $\caly_{lm}$ which defines its coordinate representation in \eqref{eq:position-rep}. The identities in \eqref{eq:a-actions} leads to
\begin{equation}
	\matens{n,z}{\caly_{lm}}{n,z}=\frac{n!}{(n-l)!}\frac{1}{(1+z\zbar )^{l}}\sum_{(lm)}\frac{(-1)^{n_2}(\zbar)^{m_1}z^{n_1}}{n_1!\,n_2!\,m_1!\,m_2!}=\frac{n!}{(n-l)!}\frac{1}{l!(l+m)!}e^{im\phi}P^m_l(\cos(\theta))
	\label{eq:sphericalharmonic}
\end{equation}
which is indeed proportional to $Y_{lm}(\theta,\phi)$. \\

We will also require knowledge about the asymptotic behaviour of the radial wave functions introduced above. To this end we first apply identities 15.3.21 and 15.4.6 from \cite{abramowitz_handbook_1964} to $\gbar_{J,l}(n,\kappa)$ and $\gbar_{Y,l}(n,\kappa)$ in order to obtain equivalent representations of these functions in terms of Jacobi polynomials. The large-$n$ behaviour, which corresponds to large radial distances, then follows from theorem 8.21.8 in \cite{szego_orthogonal_1939} as
\begin{equation}
	\gbar_{J,l}(n,\kappa)\approx\frac{\sin[(n-l-1)\kappab-l\pi/2]}{n^{l+1}}\hs{1}{\rm and}\hs{1}\gbar_{Y,l}(n,\kappa)\approx-\frac{\cos[(n-l-1)\kappab-l\pi/2]}{n^{l+1}}.
	\label{eq:gbarasymptotics1}
\end{equation}
In particular, the analogue of the Hankel function is an outgoing radial wave:
\begin{equation}
	\gbar_{H,l}(n,\kappa)=\gbar_{J,l}(n,\kappa)+i\gbar_{Y,l}(n,\kappa)\approx\frac{e^{i(n+l+1)\kappab}}{(in)^{l+1}}.
	\label{eq:gbarasymptotics2}
\end{equation}
Note that, compared to the commutative case, these expressions contain an extra radial power of $n^{-l}$. However, this is compensated for by the factor of $n!/(n-l)!\sim n^l$ appearing in \eqref{eq:sphericalharmonic}.
\subsection{Probability Currents}
We can now combine the results of the previous two sections and calculate the probability currents associated with the two important classes of states. These results will be key ingredients in the scattering formalism that follows. We start with the plane-wave given in \eqref{eq:kplanewave} and calculate the matrix elements $M_\alpha=\matens{n+1,z}{\psih^\dag\adop{\alpha}\psih\aop{\alpha}}{n+1,z}$ which define the current in \eqref{eq:current-components}. A simple calculation yields
\begin{equation}
	M_1=(n+1)\cos^2\left(\frac{\theta}{2}\right)e^{-i\kappab}\hs{1}{\rm and}\hs{1}M_2=(n+1)\sin^2\left(\frac{\theta}{2}\right)e^{+i\kappab}.
\end{equation}
Inserting this into \eqref{eq:kplanewave} and transforming to Cartesian components yields a current purely in the $z$-direction of
\begin{equation}
	j_z=\frac{\hbar\sin(\kappab)}{m_0}.
	\label{eq:planewavecurent}
\end{equation}
Next we consider a general superposition of $m=0$ partial waves of the form
\begin{equation}
	\psih=\sum_{l=0} a_l\gbar_l(\nhat-l)(l!)^2\caly_{l0}.
	\label{eq:generalpsiexpansion}
\end{equation}
Here we are interested in the probability current in the limit of large radial distances $r=\ncp(n+1)$. We find for the matrix elements
\begin{equation}
M_\alpha=\sum_{l,l'}a^*_la_{l'}\gbar^*_l(n+1-l)\gbar_{l'}(n-l')(l!)^2(l'!)^2\matens{n+1,z}{\caly_{l0}}{n+1,z}\ast_{n+1}\matens{n+1,z}{\adop{\alpha}\caly_{l'0}\aop{\alpha}}{n+1,z}
\label{eq:radialwavecurrent1}
\end{equation}
where the star product was used as in \eqref{eq:starproductfactorise} to factorise the symbol. Proceeding with this analysis requires some insight into the asymptotic behaviour of the star product itself. In \cite{kriel_eigenvalue_2012} it is shown how $\ast_n$ can be expressed as a $1/n$ expansion of the form 
\begin{equation}
\ast_n=1+\sum_{k=1}^\infty\frac{2^k}{n^k}\sum_{p,q=1}^k\overleftarrow{\partial}^p_z\Lambda^{(k)}_{p,q}(z,\zbar)\overrightarrow{\partial}^q_{\zbar}
\label{eq: starexpand}
\end{equation}
with $\Lambda^{(k)}_{p,q}(z,\zbar )$ a function of $z$ and $\zbar$. We see that as the power of $1/n$ in each term grows, so does the maximum possible order of the derivatives acting on the matrix elements standing to the left and right. In the large-$n$ limit all the terms containing negative powers of $n$ can therefore be neglected, \emph{provided that the derivatives do not bring about additional powers of $n$ through their action on the matrix elements.} From the coordinate representation of the $\caly_{lm}$ in \eqref{eq:sphericalharmonic} it is clear that the powers to which $z$ and $\zbar$ appear in the two matrix elements are set by $l$ and $l'$. A sufficient condition for the asymptotic approximation $\ast_n\approx 1$ to hold is therefore that only a finite number of the terms in the partial wave expansion of $\psih$ in \eqref{eq:generalpsiexpansion} are non-zero, i.e. that there is a maximum $l$ beyond which $a_l=0$. Of course, for a general state $\psih$ there is no reason why this condition should hold. However, as we will show in the next section, precisely this condition appears as a kinematic constraint on the scattering by a finite range potential. In that setting it will therefore be permissible to neglect the star-product in \eqref{eq:radialwavecurrent1} and so here we will proceed under this assumption. Using \eqref{eq:a-actions} to simplify the second matrix element it is found that $M_1=z\zbar M_2$ and from the form of the current in \eqref{eq:current-components} it is clear that only the radial component will be non-zero. The final form of the latter reads
\begin{equation}
	j_r(n,\theta,\phi)=\frac{\hbar}{m_0}{\rm Im}\sum_{l,l'}n^{l+l'}a_l a^*_{l'} P_l(\theta)P_{l'}(\theta)\gbar_l(n-l+1)\gbar^*_{l'}(n-l').
	\label{eq:finaljr}
\end{equation}
\section{Scattering Theory}
\label{sec:scattering-formalism}
We now turn to the main focus of this paper and consider the scattering of a particle by a finite range radial potential $V(\hat{r})$ in non-commutative space. Following the time-independent description we seek energy eigenstates of the form
\begin{equation}
	\psih=e^{i\bar{k}\hat{x}_3}+\psih^{(+)}
\end{equation}
with $e^{i\bar{k}\hat{x}_3}$ the incident plane wave and $\psih^{(+)}$ the scattered wave. Beyond the range of the potential $\psih^{(+)}$ should reduce to a solution of the free particle problem with a probability current directed away from the origin. This requirement, together with the asymptotic expression for $\gbar_{H,l}(n,\kappa)$ in \eqref{eq:gbarasymptotics2} suggests that $\psih^{(+)}$ has the form
\begin{equation}
	\psih^{(+)}=\sum_l a_l(l!)^2\psih_{H,l0}=\sum_l a_l\gbar_{H,l}(\nhat-l,\kappa)(l!)^2\caly_{l0}
\end{equation}
where axial symmetry has excluded $m\neq0$ terms from appearing in the expansion. To derive the differential cross section we require the probability currents associated with the two components of $\psih$. For the plane-wave the result is already given in \eqref{eq:planewavecurent}. For the scattered wave we can insert the expression for $\gbar_{H,l}(n,\kappa)$ in \eqref{eq:gbarasymptotics2} into that for $j_r$ in \eqref{eq:finaljr}. Together, this yields the differential cross-section as
\begin{equation}
	\frac{d\sigma}{d\Omega}=\frac{|j_r|r^2}{|j_z|}=\ncp^2\left|\sum_l a_l P_l(\cos(\theta))e^{-il\pi/2}\right|^2
	\label{eq:crosssection1}
\end{equation}
Here it is necessary to justify the use of \eqref{eq:finaljr}, which is only applicable if there is a finite number of partial waves appearing in the expansion for $\psih^{(+)}$. Consider a potential with range $R=\ncp(M+1)$ with $M\geq 0$ an integer. In the commutative case, at a fixed incident energy, the contribution of the various $l$-channels to the cross-section is generally expected to decrease with increasing $l$. Qualitatively, those channels for which $l\leq kR$ will contribute strongly, while scattering in the higher angular momentum channels will be weak but generally non-zero. In the non-commutative case there is an additional criteria for scattering to occur which is \emph{independent} of the incident energy. This is seen by analysing the potential energy term $V(\nhat)\psih$ in the Schr\"odinger equation. When $\psih$ is expanded in partial waves the form of $\caly_{lm}$ in \eqref{eq:standard-eigenfunctions2} implies that the potential will always appear in the combination $V(\nhat)(\adop{1})^{m_1}(\adop{2})^{m_2}(\cdots)$
with $m_1+m_2=l$. This implies that only the values of $V(n)$ for $n\geq l$ can play a role in determining the scattering in the $l$-channel. For a finite range of $R=\ncp(M+1)$ it holds that $V(n)=0$ for $n\geq M+1$, and therefore if $l\geq M+1=R/\ncp$ the potential is \emph{completely absent} from the $l$-channel and therefore $a_l=0$ in the expansion of the scattered wave $\psih^{(+)}$.\\

In order to determine the $a_l$ coefficients we will need to perform partial wave expansions of $\psih$ (as a general energy eigenstate) and of the incoming plane wave. The latter should mirror the well-known expansion of the plane-wave in spherical harmonics and Bessel functions. On the operator level we therefore expect that
\begin{equation}
	e^{i\bar{k}\hat{x}_3}=\sum_l d_l(l!)^2\psih_{J,l0}=\sum_{l}d_l\gbar_{J,l}(\nhat-l,\kappa)(l!)^2\caly_{l0}.
	\label{eq:planewaveexpansion}
\end{equation}
The $d_l$ coefficients can be found by switching to a representation in terms of the coherent state symbols. For the plane-wave the identities in \eqref{eq:a-actions} lead to
\begin{equation}
	\matens{n,z}{e^{i\bar{k}\hat{x}_3}}{n,z}=[\cos(\kappab)+i\cos(\theta)\sin(\kappab)]^n,
\end{equation}
and since $\matens{n,z}{\caly_{lm}}{n,z}\sim P_l(\cos(\theta))$ the problem becomes one of expanding the polynomial $(z-x)^n$ as a series of Legendre polynomials $P_l(x)$. For this we use the orthogonality of the Legendre polynomials together with identity 7.228 of \cite{gradshteyn_table_2014} to find
\begin{equation}
	d_l=i^l(2l+1)\csc(\kappab).
\end{equation}
Finally, the full wave function $\psih$ must be an asymptotic solution to the free particle problem and therefore of the form
\begin{equation}
	\psih=\sum_l(A_l\psih_{J,l0}+B_l\psih_{Y,l0}).
\end{equation}
Equating this expansion to the sum of the expansions for the plane and scattered waves reveals that
\begin{equation}
	a_l=\frac{d_l B_l}{iA_l-B_l}=id_le^{i\delta_l}\sin(\delta_l)\hs{1}{\rm with}\hs{1}\tan(\delta_l)=-\frac{B_l}{A_l}.
\end{equation}
The final expression for the differential cross-section is
\begin{equation}
	\frac{d\sigma}{d\Omega}=\frac{\ncp^2}{\sin^2(\kappab)}\left|\sum_l (2l+1)\sin(\delta_l)e^{i\delta_l}P_l(\cos(\theta))\right|^2
	\label{eq:general-dcs}
\end{equation}
while the total cross-section reads
\begin{equation}
	\sigma_{tot}=\sum_l\sigma_l=\frac{\ncp^2}{\sin^2(\kappab)}\sum_l 4\pi(2l+1)\sin^2(\delta_l).
	\label{eq:general-tds}
\end{equation}
The two expressions above are remarkably similar to their commutative counterparts, with the pre-factor $\frac{\ncp^2}{\sin^2(\kappab)}$ being the only obvious distinguishing feature. Indeed, at low incident energies we have $\bar{\kappa}\ll1$ and $\sin(\bar{\kappa})\approx k\ncp$, in which case \eqref{eq:general-dcs} and \eqref{eq:general-tds} reduce to the standard commutative results. At this stage one may therefore question whether non-commutativity will have any significant impact on the scattering at all. In the next section we show that this is indeed the case, and that strong deviations from commutative scattering can enter on the dynamical level via the phase shifts. 
\section{Spherical Well}
\begin{figure}[t]
\begin{center}
\begin{tabular}{cc}
\includegraphics[width=0.49\textwidth]{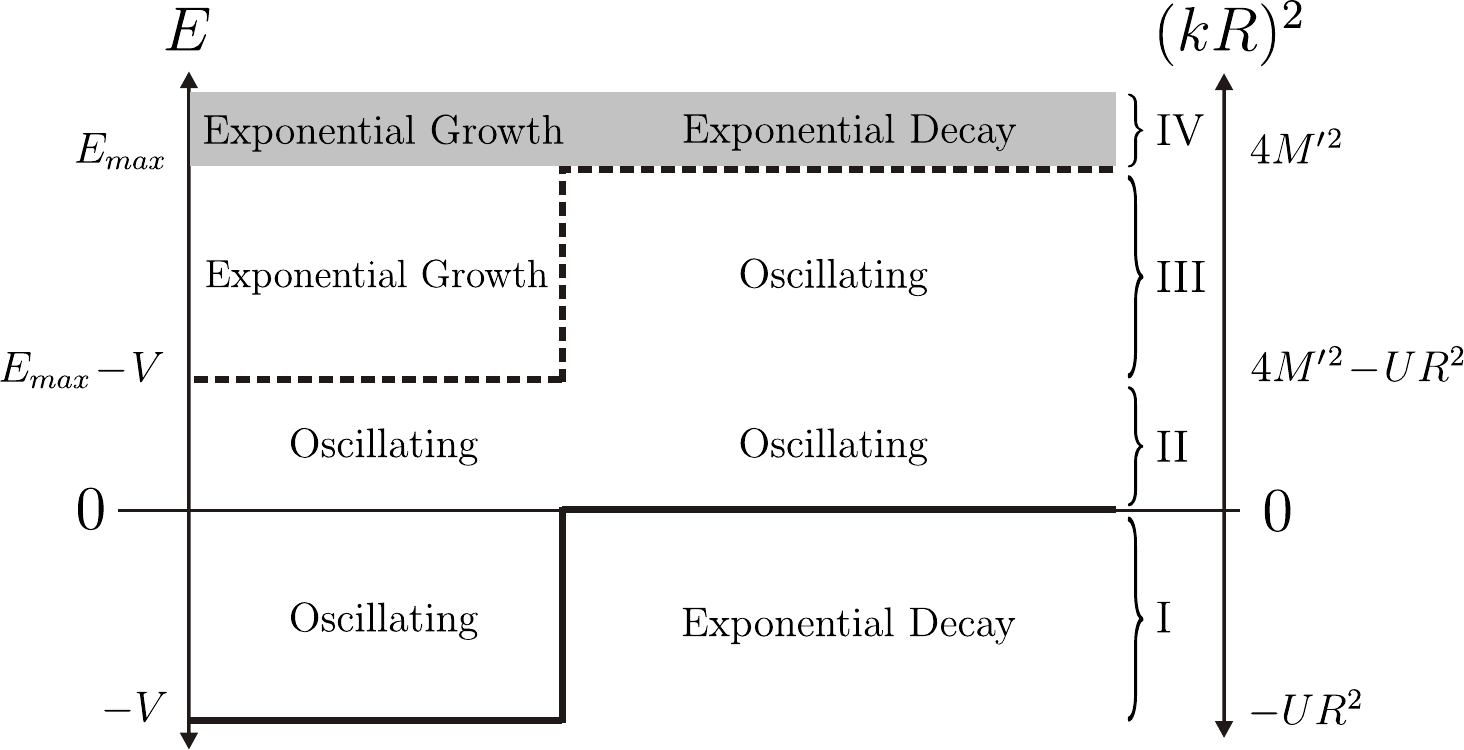}&\includegraphics[width=0.49\textwidth]{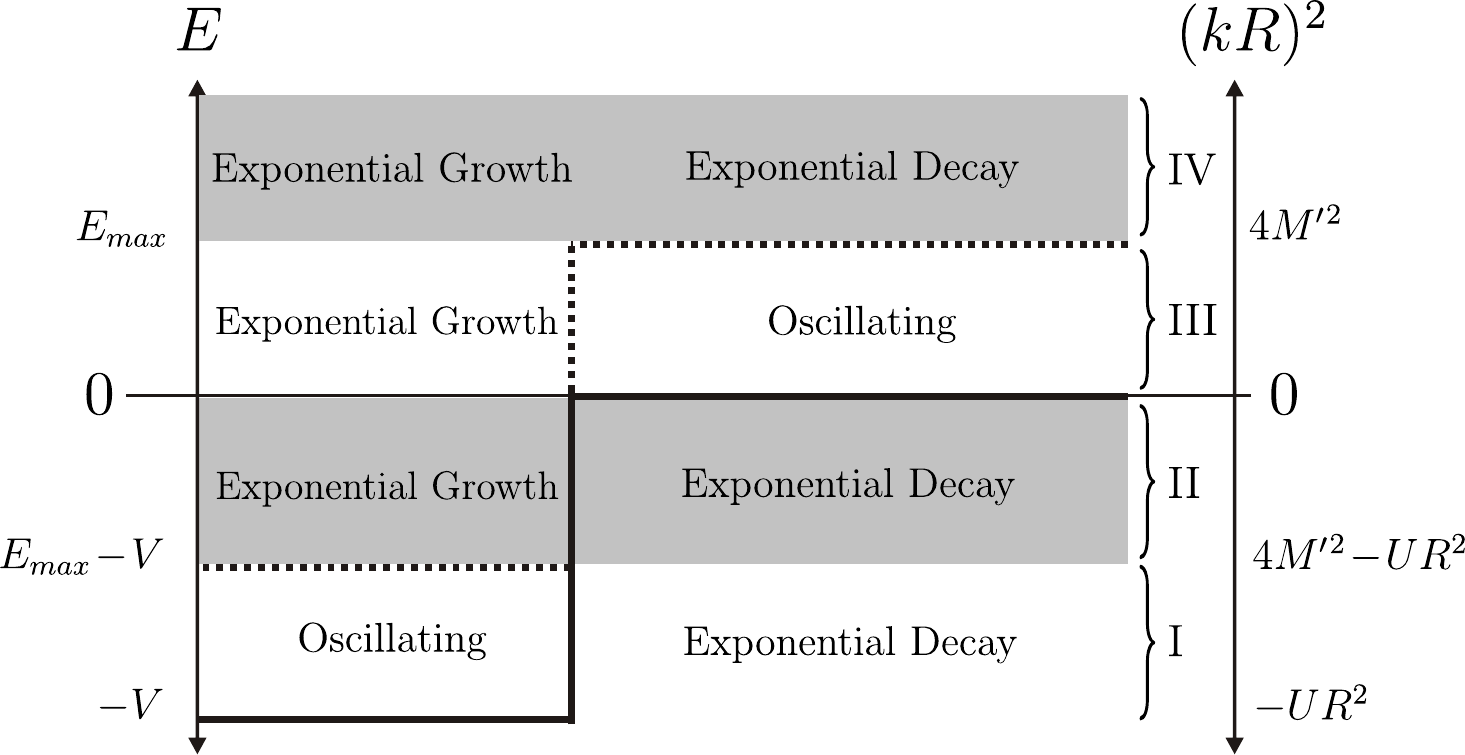}\\
(a) & (b)
\end{tabular}
\end{center}
\caption{Illustrations of the various energy regions and the corresponding behaviour of the wave function as discussed in the text. The gray regions do not support bound or scattering states. The second, dimensionless energy scale appearing on the right will be used for the scattering analysis. We use the shorthand $M'\equiv M+1$. (a) and (b) correspond to the shallow and deep well scenarios respectively.}
\label{fig:wellplot}
\end{figure}
As an application we will consider scattering off a spherical well potential. The derivation, and sensible interpretation, of the scattering results do however require some insight into the spectrum and eigenstates of this potential. Subsection \ref{sec:well-spectrum} will provide the necessary background in this regard, following the approach of \cite{chandra_spectrum_2014}. The results of the scattering analysis then follow in \ref{sec:well-scattering}.
\subsection{Spectrum and Eigenstates}
\label{sec:well-spectrum}
The Schr\"{o}dinger equation for the spherical, attractive well with radius $R=\ncp(M+1)$ reads \cite{chandra_spectrum_2014}
\begin{equation}
	-\frac{\hbar^2}{2m_0}\hat{\Delta}_\ncp\psih_{jm}-V\hat{P}\psih_{lm} = E\psih_{lm}\hs{1}{\rm with}\hs{1}V=\frac{\hbar^2U}{2m_0}
	\label{eq:hamiltonian-piecewise}
\end{equation}
and where $\hat{P} = \sum_{n=0}^{M}\sum_{k=0}^n \ket{k,\,n-k}\bra{k,\,n-k}$ projects onto configuration space states with total particle number $n\leq M$, i.e. states that are localised at radii $r\leq R$. The parameter $U$ controls the well depth while the energy is given by $E=\frac{\hbar^2\kappa^2}{2m_0\ncp}$ in the notation of \eqref{eq:planewaveenergy}. Inserting the general form of $\psih_{lm}$ from \eqref{eq:standard-eigenfunctions} into the Schr\"{o}dinger equation yields a difference equation for $\gbar(n)$ of which the (unnormalised) solution is
\begin{equation}
	\gbar_{l}(n,\kappa)=\left\{\matcc{\gbar_{J,l}(n,\kappa_i)&0\leq n \leq\mathcal{M}+1}{A_l\gbar_{J,l}(n,\kappa)+B_l\gbar_{Y,l}(n,\kappa)\hs{0.5}\mbox{}&\mathcal{M}\leq n}\right.
	\label{eq:gbar-piecewise}
\end{equation}
where $\mathcal{M}=M-l$ and $\kappa_i=\sqrt{\kappa^2+\ncp^2 U}$. The overlap of the interior and exterior solutions at $n=\mathcal{M},\mathcal{M}+1$ yields the matching conditions which determine the coefficients $A_l$ and $B_l$. Note that $\mathcal{M}$ is effectively the range of the potential in the $l$-channel, and that when $\mathcal{M}<0$ the potential disappears entirely. This forces $B_l=0$, and therefore $a_l=0$ in the expansion of the scattered wave. This is again the mechanism discussed after equation \eqref{eq:crosssection1} which limits the number of channels which undergo scattering. It is interesting to note that for a well of minimum size with $M=0$ and $R=\ncp$, only s-wave scattering can occur. Physically, this appears to suggest that such a scatterer has no structure to probe, leading to an isotropic cross-section for all incident energies.\\

Let us now consider the spectrum in more detail. Central to this discussion is the behaviour of the functions $\gbar_{J,l}(n,\kappa)$ and $\gbar_{Y,l}(n,\kappa)$ in different energy regimes. For kinetic energies between $0$ and $E_{max}$, so for $\kappa\in[0,2]$, both these functions display the usual oscillatory behaviour. At negative kinetic energies, i.e. $\kappa^2<0$, both grow exponentially, although the combination $\gbar_{J,l}(n,\kappa)+i\gbar_{Y,l}(n,\kappa)$ will be exponentially decaying. The same situation results for kinetic energies which exceed $E_{max}$, so for $\kappa>2$, although here it is $\gbar_{J,l}(n,\kappa)-i\gbar_{Y,l}(n,\kappa)$ that will decay exponentially. Depending on the depth of the well and the incident energy we can have different combinations of these scenarios in the interior and exterior regions. The spectrum is of course determined by whether it is possible to match these solutions at $n=\mathcal{M},\mathcal{M}+1$. We consider two cases:
\begin{itemize}
	\item \textbf{A Shallow Well with $V<E_{max}$:} See figure \ref{fig:wellplot}(a). Here region II contains a continuum of scattering states for which the particle's kinetic energy inside the well does not exceed $E_{max}$. At higher incident energies in region III the kinetic energy inside the well will exceed $E_{max}$ and the interior solution then grows exponentially in magnitude with increasing $n$. This solution can still be matched with the oscillatory exterior solution, and so we also find a continuum of scattering states in this region. Note that here the behaviour inside the well is more akin to what one would expect for a \emph{repulsive} potential. This will also be reflected in the scattering results. In region IV it is impossible to match the growing and decaying solutions, resulting in an empty spectrum. Region I will contain the usual bound states.
	
	\item \textbf{A Deep Well with $V>E_{max}$:} See figure \ref{fig:wellplot}(b). Here the particle's kinetic energy inside the well exceeds $E_{max}$ for all the scattering states in region III. The exponential behaviour in the interior suggests that the well will act as an effective repulsive barrier for all incident energies. In region II no bound states will be found, since it is impossible to satisfy the matching conditions here. This illustrates another important result: \emph{the well can only support a finite number of bound states, regardless of its depth.} It can be shown that in the $l$-channel a maximum of $M-l+1$ bound states are supported \cite{chandra_spectrum_2014}. These appear in region I and are separated from the scattering states by an energy gap of $V-E_{max}$. 
\end{itemize}
\subsection{Scattering}
\label{sec:well-scattering}
The phase shifts are the basic input for the scattering formalism. These are determined by the ratio of the $A_l$ and $B_l$ coefficients in \eqref{eq:gbar-piecewise}, which in turn are fixed by the matching conditions at $n=\mathcal{M},\mathcal{M}+1$. We find that
\begin{equation}
	\frac{B_\ell}{A_\ell}=\frac{\gbar_{J,\ell}(\mathcal{M}+1,\kappa)\,\gbar_{J,\ell}(\mathcal{M},\kappa_i)-\gbar_{J,\ell}(\mathcal{M},\kappa)\,\gbar_{J,\ell}(\mathcal{M}+1,\kappa_i)}{\gbar_{J,\ell}(\mathcal{M}+1,\kappa_i)\,\gbar_{Y,\ell}(\mathcal{M},\kappa)-\gbar_{J,\ell}(\mathcal{M},\kappa_i)\,\gbar_{Y,\ell}(\mathcal{M}+1,\kappa)}
\label{eq:ab-ratio}
\end{equation}
in which the various energy related parameters are defined by
\begin{equation}
	E=\frac{\hbar^2\kappa^2}{2m_0\ncp^2}=\frac{\hbar^2k^2}{2m_0}\hs{2}\kappa=2\sin\left(\frac{\kappab}{2}\right)\hs{2}\kappa=k\ncp\hs{2}\kappa_i=\sqrt{\kappa^2+\ncp^2 U}.
\end{equation}
For presenting the scattering results we will use the three dimensionless parameters $kR$, $UR^2$ and $M=R/\ncp-1$ which provide, in a rough sense, relative measures of the incident energy, the strength of the potential, and the non-commutative length scale. In particular, $M\rightarrow\infty$ corresponds to the commutative limit.\\

\begin{figure}[ht]
\begin{center}
\begin{tabular}{cc}
\includegraphics[width=0.48\textwidth]{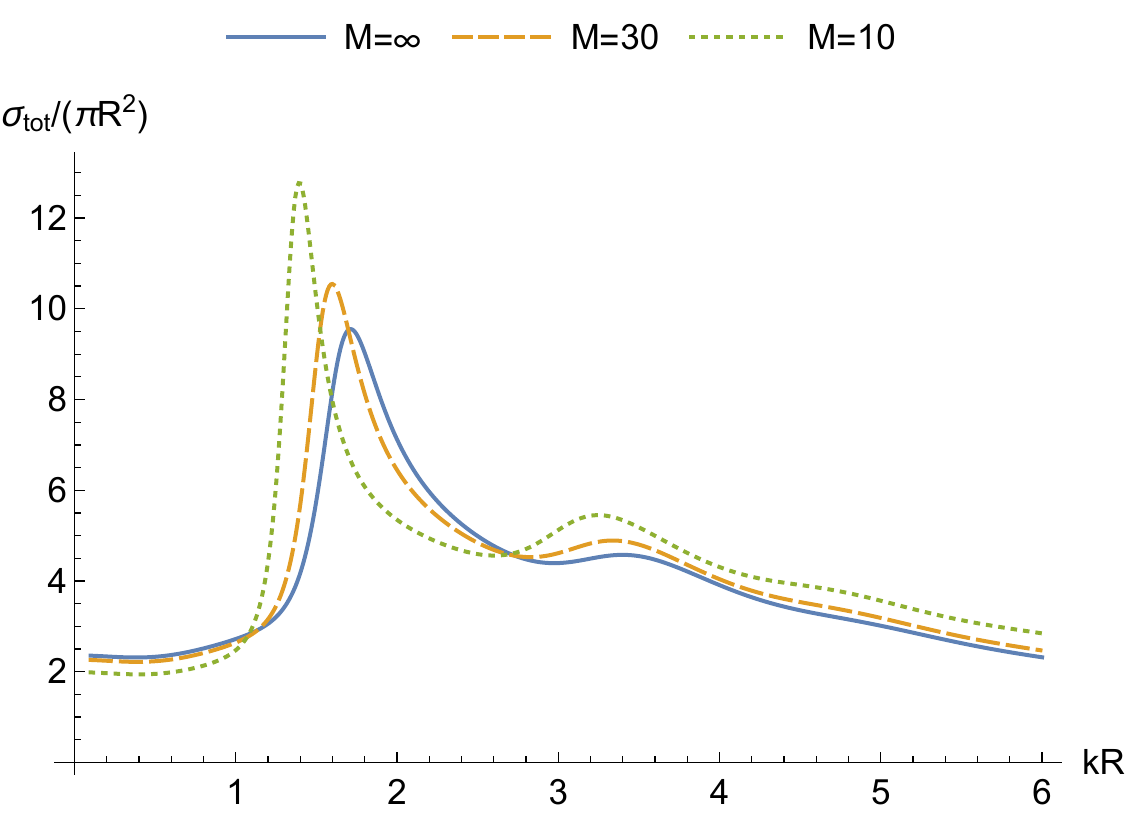} \hs{0.2}& \includegraphics[width=0.48\textwidth]{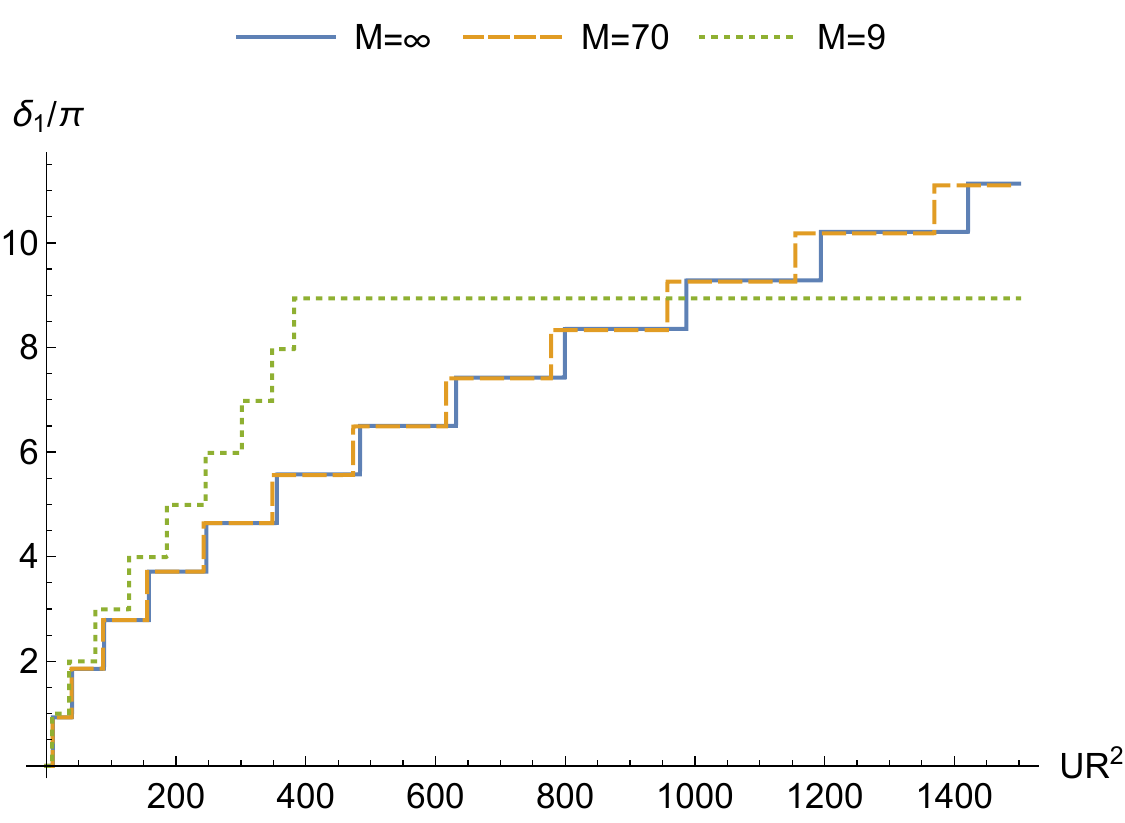}\\
(a) & (b)\\
\includegraphics[width=0.48\textwidth]{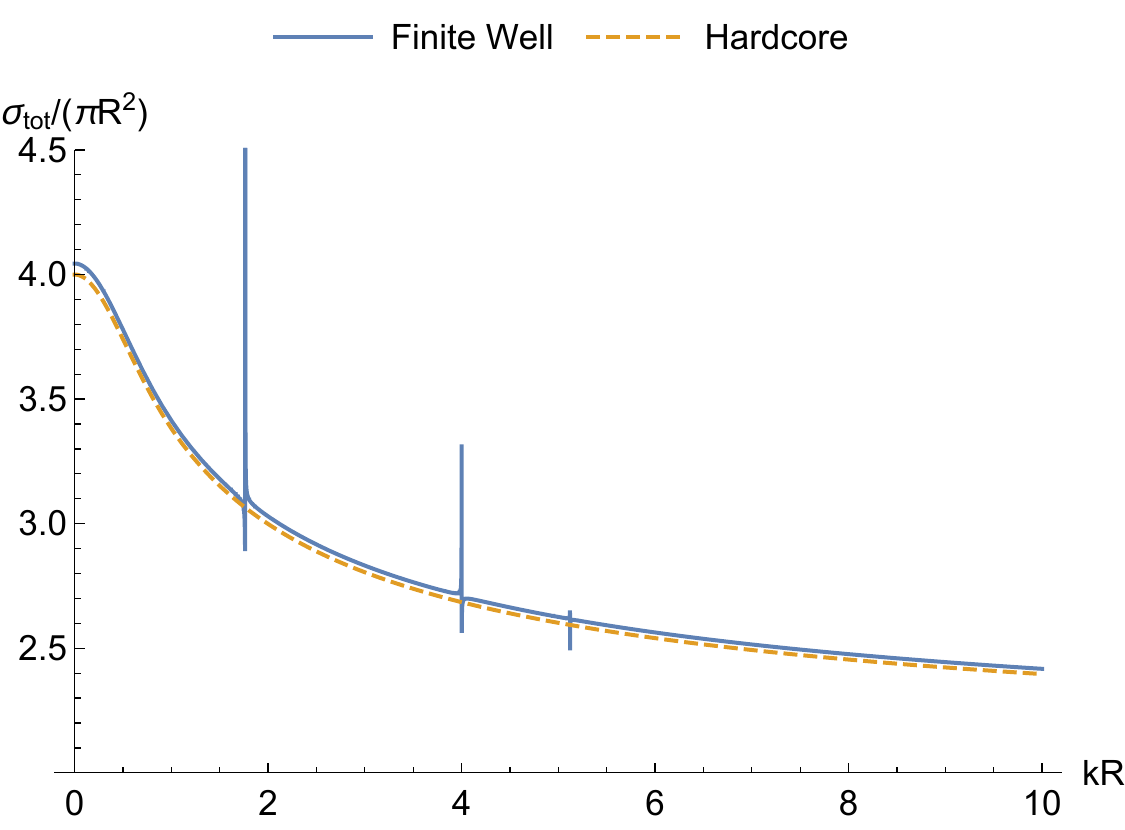} & \includegraphics[width=0.48\textwidth]{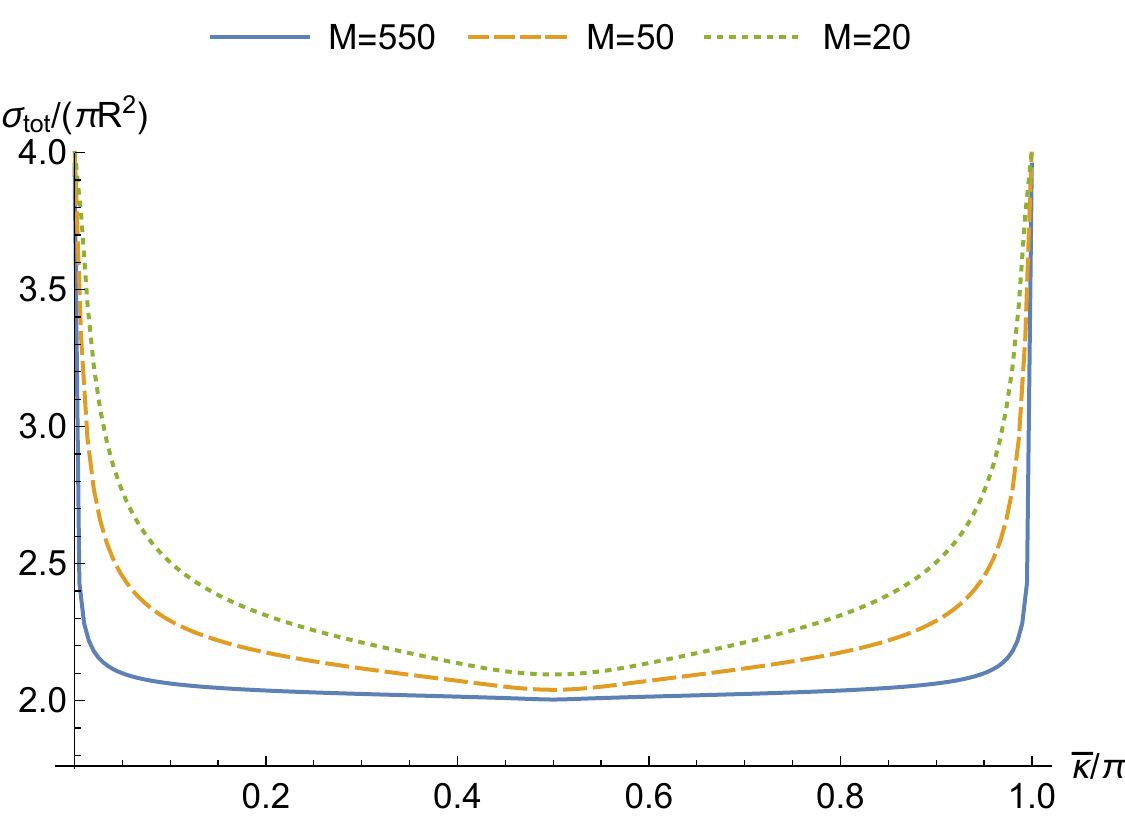}\\
(c) & (d)
\end{tabular}
\end{center}
\caption{See discussion in text. (a) The total cross-section as a function of incident energy with $UR^2=15$. (b) The phase-shift of the $l=1$ channel as a function of the well-depth at $kR=0.1$. (c) The total cross-section as a function of the incident energy for a finite attractive well and a repulsive hardcore potential. Here $M=100$ and $UR^2=40768$. (d) The total cross-section as a function of the incident energy for a repulsive hardcore potential.}
\label{fig:scattering-plots}
\end{figure}

In figure \ref{fig:scattering-plots}(a) we show the total cross-section as a function of the incident energy for three different values of $M$ and with $UR^2=15$. As $M$ decreases the non-commutative length grows and the peaks in the cross-section become sharper and move to lower energies. There appears to be a general trend that non-commutativity results in the poles of the S-matrix moving closer to the real energy axis resulting in sharper, longer lived resonances.\\
Figure \ref{fig:scattering-plots}(b) shows the phase shift in the $l=1$ channel as a function of $UR^2$ at a low incident energy $kR=0.1$ for different values of $M$. We see that as the well-depth grows the phase shift increases in sharp steps of $\pi$. These are the zero-energy resonances corresponding to the formation of bound states in the well. This pattern continues indefinitely in the commutative case, but for finite $M$ the phase shift eventually saturates at a constant value. For $M=9$ this occurs at a value of $\delta_1\approx 9\pi$, which reflects the maximum number of $M-l+1=9$ bound states the well can accommodate in this channel. This is essentially the content of Levinson's theorem, demonstrated here in the non-commutative setting. Furthermore, $\delta_1\approx 9\pi$ is also the phase shift of the hardcore \emph{repulsive} well. This reflects the fact that when we cross from a shallow to a deep well, as per the discussion in section \ref{sec:well-spectrum}, the formation of bound states cease and the well effectively becomes repulsive. This effect also appears at a fixed well-depth when the incident energy is increased across the boundary between regions II and III in figure \ref{fig:wellplot}(a). This is illustrated in figure \ref{fig:scattering-plots}(c) in which $M=100$ and $UR^2$ is chosen such that $4(M+1)^2-UR^2=6^2$. The boundary between the two regions therefore occurs at $kR=6$. In region II the cross-section is seen to still exhibit some structure in the form of three extremely sharp resonances. However, when $kR$ crosses into region III the well becomes effectively repulsive, and the cross-section exhibits the same structureless trend as that of the hardcore repulsive potential.\\

Let us now consider hardcore scattering in more detail. As suggested by the preceding discussion we expect this scenario to result from \emph{both} the $V\rightarrow\infty$ and $V\rightarrow-\infty$ limits. Indeed, it can be verified that in both cases \eqref{eq:ab-ratio} reduces to
\begin{equation}
	R_l=\frac{B_l}{A_l}=-\frac{\gbar_{J,l}(\mathcal{M},\kappa)}{\gbar_{Y,l}(\mathcal{M},\kappa)}.
\end{equation}
Another interesting feature of this scenario is that the scattering exhibits a kind of high-low energy duality reflected by the fact that $R_l(\kappab)=-R_l(\pi-\kappab)$. In terms of the incident energy this implies that the scattering cross-section at $E$ and $E_{max}-E$ will be identical. Figure \ref{fig:wellplot}(d) shows the total cross-section as a function of $\kappab$ for different values of $M$. The symmetry of the cross-section around $\kappab=\pi/2$, i.e. $E=E_{max}/2$, is clearly visible. Even at finite $M$ the cross-section tends to $\sigma_{tot}=4\pi R^2$ at low energy, which is exactly the well-known commutative result. As the energy increases the cross-section deceases, reaching a minimum value at $\kappab=\pi/2$ which converges to the commutative high-energy result of $\sigma_{tot}=2\pi R^2$ with increasing $M$.
\section{Summary and conclusions}
\label{sec:conclusions}
We have developed scattering theory within non-commutative fuzzy space. This required adopting POVM-based definitions of the spacial probability density and associated current. Expressions relating the scattering cross-section to the phase shifts were derived and found to be remarkably similar to their commutative counterparts; presumably a consequence of fundamental kinematic restrictions and the rotational invariance of the coordinate algebra. More pronounced deviations from commutative behaviour were shown to enter via the phase-shifts, which depend sensitively on how the non-commutative length parameter compares to the other length-scales in the problem. It was found that for a potential with range $R$ only the first $R/\ncp$ angular momentum channels will undergo scattering, thereby limiting the resolution on which features of the scatterer can be resolved. This underlines the notion of $\ncp$ as being a fundamental minimum length-scale beyond which the scattering potential cannot be probed. A second interplay exists between $\ncp$ and the projectile's wavelength which results in the normally oscillatory free particle wave-functions exhibiting exponential behaviour at very high kinetic energies. This is the origin of the cut-off in the incident energy and the fact that a sufficiently deep attractive well can adopt features of a repulsive potential. These effects are present even at low incident energies. We also observed what appears to be a general trend that non-commutativity results in scattering resonances shifting to lower energies and becoming sharper and longer lived.
\bibliography{scattering}
\end{document}